# Cloud Computation and Google Earth Visualization of Heat/Cold Waves: A Nonanticipative Long-Range Forecasting Case Study

Zubov Dmytro
Tecnológico de Monterrey, School of Engineering and Sciences, Campus San Luis Potosí
Av. Eugenio Garza Sada 300, Lomas del Tecnologico, 78211 San Luis Potosi, S.L.P, Mexico
dzubov@ieee.org

**Abstract.** Long-range forecasting of heat/cold waves is a topical issue nowadays. High computational complexity of numerical and statistical models' design is a bottleneck for the forecast process. In this work, Windows Server 2012 R2 virtual machines are used as a high-performance tool for the speed-up of the computational process. Six D-series and one standard tier A-series virtual machines were hosted in Microsoft Azure public cloud for this purpose. Visualization of the forecasted data is based on the Google Earth Pro virtual globe in ASP.NET web-site against http://gearth.azurewebsites.net (prototype), where KMZ file represents geographic placemarks. The long-range predictions of the heat/cold waves are computed for several specifically located places based on nonanticipative analog algorithm. The arguments of forecast models are datasets from around the world, which reflects the concept of teleconnections. This methodology does not require the probability distribution to design the forecast models and/or calculate the predictions. Heat weaves at Annaba (Algeria) are discussed in detail. Up to 36.4% of heat waves are specifically predicted. Up to 33.3% of cold waves are specifically predicted for other four locations around the world. The proposed approach is 100% accurate if the signs of predicted and actual values are compared according to climatological baseline. These high-accuracy predictions were achieved due to the interdisciplinary approach, but advanced computer science techniques, public cloud computing and Google Earth Pro virtual globe mainly, form the major part of the work.

## 1. Introduction

Numerical and statistical methods of weather prediction are characterized by high computational complexity because design of the forecast models has many solutions to a problem. Moreover, some particular cases of nonstationary time series (e.g. atmospheric air temperature data) do not have enough of the right data. Nowadays, numerical and synoptic methods are applied for the short and medium-range forecasting mainly because of low forecast accuracy at long term and use of highly complex equations.

Heterogeneous algorithms are applied for the long-range weather forecasting – seasonal time series (Cheng Qian et al. 2011; Qiang Song 2011), neural networks (Ahmet Erdil & Erol Arcaklioglu 2013; Gyanesh Shrivastava et al. 2012), probability theory (Nan-Jung Hsu et al. 1999; Sadokov et al. 2011), a moving blocks bootstrap method (Feng, DelSole, & Houser 2011), ensemble forecasting (Astahova & Alferov 2008; Hailing Zhang & Zhaoxia Pu 2010), scenarios of anthropogenic forcing (Collins et al. 2013; Bardin 2011), the ENSO cycle usage (Higgins et al. 2004; Ropelewski & Halpert 1986), etc. The nonlinearity and sensitivity of the existing forecast models, possible small errors and deviations of initial conditions (downpour, dust, hailstorm, sand, pollution, etc.), random measurement errors, and lack of right data (i.e. coverage error) combine to reduce the forecast accuracy and complicate the design of models (Douglas & Englehart 2007; Fathalla A Rihan & Chris G Collier 2010; Tyndall et al. 2010). Self-organizing inductive modelling shows good results when enough of the right data that would be needed in more conservative approaches is fundamentally not obtainable. They have already been applied to long-range weather forecasting (Madala & Ivakhnenko 1994). Successful applications of inductive modeling have been realized in other fields as well (e.g. stock market, economic systems, noise immunity, decision trees, data mining and neural networks).

A wide range of meteorological variables (e.g. air pressure and temperature, water vapor, precipitation, wind, snow depth) at different locations is used for the forecasting (Kattsov 2010). They interact constantly, and some variables may be evaluated using the others in accordance with known teleconnection patterns (Nada Pavlovic Berdon 2013). Thus, the design of forecast models must involve the full set of meteorological variables. However, temperature and precipitation are the targets of the long-range forecasting, mainly because of practical needs. The precipitation and the air temperature data are related to each other (Van Den Dool & Nap 1985). Correlation analysis shows that the precipitation forecasting is effective over a period of two weeks, air temperature over a much longer period (Zubov & Vlasov 2004). The impact is increased further because extremes can be used for the correction of forecasted averages (Zubov 2013).

In (Zubov et al. 2015), a nonanticipative forecast algorithm is applied for the long-range prediction of heat/cold waves. The forecast model's arguments are daily mean air temperature, mean visibility, wind speed, and dew point, as well as maximum and minimum temperatures reported during the day from 66 places around the world. In addition, sea level, sea level pressure, sea surface temperature, SOI, equatorial SOI, and multivariate ENSO index are used. An objective function selects datasets from two locations with appropriate lead-time and summation interval, which have maximum (or minimum) sum compared with the rest of data fourfold at least for the extreme events occurred on the learning sample. Up to 18.2% of all extremes were specifically predicted for the Ronald Reagan Washington National Airport. The methodology has 100% accuracy with respect to the sign of predicted and actual values. It is more useful than current methods for predicting heat/cold waves because this approach does not require the probability distribution to design the forecast models and/or calculate the predictions. However, only one location (Ronald Reagan Washington National Airport) was discussed. In addition, 66 places do not form a representative sample. In (Zubov 2015), 119 places were used for the design of the forecast model. Here, the forecast models are designed in the high-performance Microsoft Azure virtual machines (VMs). Nevertheless, only one location was discussed for prediction. Also, data visualization was not presented.

Reasoning of the statistical forecast models takes a lot of time – two weeks for one place approximately (Zubov 2015). Hence, high-performance computing (HPC) has to be in use. It can be classified into three main categories nowadays – supercomputers, cloud computing, and powerful desktop computers. Tianhe-2 is an example of supercomputers. In fact, this is fastest (Jun 2015) supercomputer system with a theoretical peak performance of the system of 54.9024 petaflops (Rezaur Rahman 2013). The adiabatic quantum computer D-Wave Two (Zhengbing Bian, Fabian Chudak et al. 2014) is an example of non-traditional supercomputer too which can use a representation of qubits as historical data. The latest version D-Wave 2X with more than 1000 qubits was released in Aug 2015. Here, the design of the forecast model is based on the quantum annealing for minimization of the Ising model's dimensionless energy. However, access to supercomputers is restricted because of high use value mainly. Microsoft Azure, Amazon Web Services, and Google Cloud Services are examples of popular public clouds. In fact, Azure VMs were used mainly in this project as the high-performance tool for the design of forecast models.

The geographic data visualization is a front-end part of the project. It includes graphical representation of the maps and charts according to end-user's requirements and intelligent visual queries based on keyword method and graphical interface (Martin Dodge et al. 2008). Nowadays, Google Earth Pro is free integrated development environment, which displays the satellite images and other related info using 3D globe model of Earth. In fact, the above-discussed approaches can be implemented by Google Earth Pro virtual globe and appropriate web apps.

Based on the above-stated brief analysis of previous studies, this paper main goal is the high-performance computation and data visualization of heat/cold waves. Microsoft Azure VMs, Google Earth Pro virtual globe, and nonanticipative analog method for the long-range forecasting of heat/cold waves (several places are discussed) are in use, respectively.

This paper is organized as follows: In Section 2, a nonanticipative analog method for the long-range forecasting of heat/cold waves is discussed. In Subsection 2.1, data sources and nonanticipative analog principle for the long-range forecasting are presented. In Subsection 2.2, a long-range forecasting of heat/cold waves at different locations as well as a particular case of Annaba (Algeria) are discussed. In Section 3, the design of forecast models in Microsoft Azure VMs as well as the data visualization based on Google Earth Pro virtual globe are shown. Conclusions are summarized in Section 4.

## 2. A nonanticipative analog method for the long-range forecasting of heat/cold waves. Results and discussion

### 2.1. The data sources and nonanticipative analog principle for the long-range forecasting of heat/cold waves

It is assumed that some event (or group of events) $A(j)$ has an impact on another event $B(j')$, where $B(j')$ – extreme air temperature (i.e. heat/cold wave), $j, j'$ – event dates and $(j'-j)>0$. Here, a value is considered extreme if the difference between this value and its climatological baseline is greater than two standard deviations (SD) in absolute units. Air temperature correlation analysis (Zubov 2013) shows that enough of the right data is not obtainable. In this case, fourfold repetitions of extreme air temperature for the given lead-time $(j'-j)$ after the same event $A(j)$, on the learning sample is taken to establish the validity of $A(j)$ as a predictor. The dependency thus discovered is used as objective function for prediction. In fact, the nonanticipative forecasting methodology presented in Section 2 is similar to (Zubov 2015), but several locations are discussed here.

The heat/cold waves cannot be identified completely nowadays, which complicates the use of the weather data generators (Adelard et al. 2000). In this case, NOAA Satellite and Information Service (http://www7.ncdc.noaa.gov/CDO/cdo) is used as a main free data source since 1973 yr, providing 119 datasets from around the world: daily mean air temperature, mean visibility in miles, mean wind speed in knots, mean dew point in Fahrenheit, maximum and minimum temperatures in Fahrenheit reported during the day for the appropriate place. In addition, Darwin and Tahiti sea level pressures, southern oscillation index (SOI), equatorial SOI, sea surface temperature, multivariate ENSO index (monthly mean), daily mean sea level data at Aburatsu, Japan (GLOSS database, University of Hawaii Sea Level Center; http://ilikai.soest.hawaii.edu/woce/wocesta.html) are used. Hence, 131 datasets are taken into consideration – $X_i=\{x_{i1}, x_{i2}, ..., x_{ij}, ...\}$, $i = \overline{1,131}$, $j = \overline{1,15340}$ ($j=1$ corresponds to Jan 1, 1973, $j=15340$ – to Dec 31, 2014; some stations are presented in Table 1). These datasets and resources are selected because of free public access, daily mean air temperatures, and data archives since 1973 yr at least. Preprocessing standardizes the data using climatological baseline $\overline{x}_{ij}$ calculated as expectations for the appropriate date from Jan 1, 1973 to Dec 31, 2013 (e.g. $x_{114,4}=39.3°F$, $\overline{x}_{114,4}=37.5°F$, $x^*_{114,4}=-1.8°F$; 114 is the ordinal number of Ronald Reagan Washington National Airport):

$$x^*_{ij} = x_{ij} - \overline{x}_{ij}. \tag{1}$$

Considering the Ronald Reagan Washington National Airport dataset, the positive $x^+_{114,j_+}$ ($j_+ = \overline{1,364}$) and negative $x^-_{114,j_-}$ ($j_- = \overline{1,309}$) extremes are studied from 1973 to 2014 ($x^+_{114,j_+}, x^-_{114,j_-} \in E$, $E$ – set of extremes). The data are split into learning (from 1975 to 2010: $j = \overline{731,13879}$; years 1973 and 1974 are reserved because of lead-time $l$ and summation interval of length $n$ which are up to one year each) and validation (from 2011 to 2014: $j = \overline{13880,15340}$) samples.

Table 1. Some stations around the world, which are used for the design of forecast models.

| $i$ | Country or region | Station (named by NOAA) | $i$ | Country or region | Station (named by NOAA) |
|---|---|---|---|---|---|
| 1 | Algeria | Annaba | 113 | United Kingdom | Heathrow Airport |
| 2 | American Samoa | Tafuna-Pago International AP | 114 | USA | Ronald Reagan Washington National Airport |
| 3 | Antigua And Barbuda | V C Bird INTL | 115 | Uruguay | Carrasco INTL |
| 4 | Argentina | Ministro Pistarini INTL | 116 | Uzbekistan | Yuzhniy |
| 5 | Aruba | Reina Beatrix INTL | 117 | Vanuatu | Aneityum |
| 6 | Australia | Canberra Airport | 118 | Venezuela | Simon Bolivar INTL |
| ... | ... | ... | 119 | Vietnam | Danang INTL |

Considering the Ronald Reagan Washington National Airport dataset ($i$=114), the objective function that defines an event $A(j)$ as a precursor to an extreme event $B(j')$ is based on situations in the learning sample where

$$\sum_{k=0}^{n-1}\left(x^*_{i_1,(j'-k-l)} + x^*_{i_2,(j'-k-l)}\right)\bigg|_{\substack{i_1,i_2 \in I \\ x^*_{114,j'} \in E \\ j'=[731,13879] \\ l \in L \\ n \in N}} > \underset{i_1,i_2,n,l}{Max} \vee \sum_{k=0}^{n-1}\left(x^*_{i_1,(j'-k-l)} + x^*_{i_2,(j'-k-l)}\right)\bigg|_{\substack{i_1,i_2 \in I \\ x^*_{114,j'} \in E \\ j'=[731,13879] \\ l \in L \\ n \in N}} < \underset{i_1,i_2,n,l}{Min}, \quad (2)$$

$$\underset{i_1,i_2,n,l}{Max} \equiv \max_{\substack{j=[731,13879] \\ x^*_{114,j} \notin E}} \sum_{k=0}^{n-1}\left(x^*_{i_1,(j-k-l)} + x^*_{i_2,(j-k-l)}\right)\bigg|_{\substack{i_1=[1,131] \\ i_2=[i_1,131] \\ l=[14,365] \\ n=[1,365]}}, \quad \underset{i_1,i_2,n,l}{Min} \equiv \min_{\substack{j=[731,13879] \\ x^*_{114,j} \notin E}} \sum_{k=0}^{n-1}\left(x^*_{i_1,(j-k-l)} + x^*_{i_2,(j-k-l)}\right)\bigg|_{\substack{i_1=[1,131] \\ i_2=[i_1,131] \\ l=[14,365] \\ n=[1,365]}}$$

($k$ – temporal summation index (days); $n$ – length of summation interval (days); $l = j'-j$ – lead-time (days); $i_1, i_2 \in I$, $n \in N$, $l \in L$ for interrelated sets $I \subset [1,131]$, $N \subset [1,365]$, $L \subset [14,365]$ of meteorological variables, possible lengths of summation intervals, and lead-times, respectively; here, all variables are of integer type). Cardinality of a set $I$ equals two because of high computational complexity of the nonanticipative analog algorithm. The sets $I$, $N$, and $L$ encompass the precursor events $A(j)$ for a given extreme event $B(j')$, for $j = j' - l$. Then, input datasets $i_1$ and $i_2$ with appropriate lead-time $l$ and summation interval of length $n$ are selected to define a prediction rule if the sum of meteorological variables from datasets $i_1$ and $i_2$ is greater than maximum $Max$ (or less than minimum $Min$) fourfold at least (with a time difference greater than 30 days) on the learning sample, for cases where $x^*_{114,j'} \in E$, i.e. where there is an extreme event $B(j')$ at day $j'$.

Hence, every extreme selection rule includes six parameters – the indices of two datasets $i_1$ and $i_2$, the lead-time $l$, the summation interval of length $n$, maximum $Max$, and minimum $Min$. $Max$ and $Min$ are computed as the maximum and minimum values of the sums over the datasets $i_1$ and $i_2$, with the same summation interval of length $n$ and lead-time $l$, where an extreme event does not occur in the learning sample ($i$=114 is the index of the Ronald Reagan Washington National Airport dataset, and has to be altered for prediction of extreme events at other locations).

**2.2. A long-range forecasting of heat/cold waves at different locations using a nonanticipative analog method**

A nonanticipative analog method consists of four main steps:
1. Generation of the prediction rules.
2. Analysis of the prediction rules. Rules with time frames, which are not concentrated at the same period of the year (up to 4 months period is taken into consideration here), are excluded.
3. Generation of possible extremes.
4. Analysis of the possible extremes. The extremes with time frames, which do not correspond to the periods of the appropriate rules, are excluded.

These steps are illustrated by the forecasting of heat/cold waves at Annaba (Algeria) in detail. Results of the heat/cold waves' prediction from 2011 to 2014 at different locations (places were selected randomly) are presented in Table 2.

Table 2. Results of the heat/cold waves' prediction from 2011 to 2014 at different locations.

| $i$ | Country | Station (named by NOAA) | Forecast accuracy for the heat waves, % | Forecast accuracy for the cold waves, % |
|---|---|---|---|---|
| 1 | Algeria | Annaba | 36.4 | 0 |
| 70 | Macedonia | Skopje | 0 | 7.1 |
| 96 | Russia | Dolgoprudnyj | 0 | 7.7 |
| 112 | Ukraine | Zhulyani | 0 | 33.3 |
| 114 | USA | Ronald Reagan Washington National Airport | 26.3 | 0 |

Thirty nine prediction rules were found for the forecasting of heat waves at Annaba. First rule consists of the following parameters: $i_1$=21, $i_2$=64, $n$=2 days, $l$ =60 days, $Min$=-43.6$^0$F, $Max$=30.4$^0$F. This rule can be described by the tuple (21, 64, 2, 60, -43.6, 30.4) concisely; it was in use fourfold on learning sample on Aug 24, 1987, Apr 21, 1996, May 11, 1997, and Dec 29, 2009. Other rules are presented in Table 3 similarly. Analysis of these rules shows that some of them are not concentrated at the same time frame. Hence, they are excluded or taken into consideration as additional info. For example, the first rule is excluded because it was used in Aug, Dec, Apr, and May, which do not belong to the same time frame. Second rule is taken into consideration because it was used fourfold in May, twice in Jun, and once in Sept. In Table 3, rule numbers with sign "-" are excluded from discussion.

These rules allow prediction of heat waves as documented in Table 4. All invocations of the above rules are listed in the table. The forecasted extremes were compared to maximal or minimal observed values in the same sector of the month (with months divided into thirds), since prediction of the exact date is not to be expected. Eight heat waves at Annaba were predicted correctly – on Mar 15, Jun 18, Jul 11-13, 2011, Nov 6, 2013, Jan 19, Jul 19-20, Sept 19-20, and Nov 23 – Dec 1, 2014. For instance, the heat wave on Feb 24, 2011 was identified using the daily mean air temperatures from Lisbon (Portugal, 93) and Rarotonga International Airport (Cook Islands, 24). This place has 22 groups of heat waves from 2011 to 2014. Hence, the proposed nonanticipative method predicted 36.4% of actual heat waves (61.5% accuracy if forecasted extremes are discussed only). In addition, the nonanticipative long-range forecasting of the Annaba heat waves is 100% accurate if the signs of predicted and actual values are compared according to climatological baseline.

**3. Design of the forecast models using Microsoft Azure virtual machines. Google Earth Pro visualization of heat/cold waves**

Public cloud platform-as-a-service is a high-performance tool for NP-complex tasks (Collier & Shahan 2015). In particular, Windows Azure VMs can compute the above-discussed nonanticipative analog algorithm. The forecast software was developed using Windows Forms and Delphi integrated development environment. Hence, VMs can use standard operating system Windows Server 2008/2012 R2 to execute files. In Jul 2014, presented methodology got the Microsoft Research Climate Data Award, which allows to start Microsoft Azure public cloud with 32 processors Intel(R) Xeon(R) E5-2660 2.20 GHz. The six D-series VMs from Climate Data Award as well as one A-series standard tier VM from Windows Azure Educator grant (two Intel(R) Xeon(R) E5-2673 2.4 GHz processors) took part in the design of the forecast models – management portal is shown in Fig. 1. HPC public cloud resources were split into several VMs because of possibility to variate the payment when other computational tasks like point-to-site network with the gateway (Collier & Shahan 2015) are started and need resources. In this case, some of the VMs are stopped while others continue to work. In addition, VMs from one subscription were hosted in different datacenters because of performance – it was found that VMs work slower sometimes if they are in the same datacenter.

Visualization of the forecasted data is based on Google Earth Pro virtual globe (Kelly L. Murdock 2009). Here, KMZ file represents the geographic placemarks. In fact, KMZ is a compressed form of KML files, which allows to load data faster. Prototype of the web-site was developed using ASP.NET technology and hosted in Windows Azure public cloud against http://gearth.azurewebsites.net. Screenshot is shown in Fig. 2.

**4. Conclusions**

In this paper, the cloud computation using Windows Azure VMs and data visualization in Google Earth Pro virtual globe are discussed for the nonanticipative long-range forecasting of heat/cold waves.

Six high-performance D-series VMs and one A-series standard tier VM with Windows Server 2012 R2 operating system were used for the design of forecast models. VMs were hosted in the Windows Azure public cloud.

Visualization of the forecasted data is based on Google Earth Pro virtual globe in ASP.NET web-site. KMZ file represents the geographic placemarks. Prototype of the web-site was developed using ASP.NET technology and hosted in Windows Azure public cloud against http://gearth.azurewebsites.net.

Table 3. The prediction rules for the forecasting of Annaba heat waves.

| Rule No. | Rule tuple (forecasted actual extremes) | Rule No. | Rule tuple (forecasted actual extremes) | Rule No. | Rule tuple (forecasted actual extremes) |
|---|---|---|---|---|---|
| 2 + | 28, 18, 6, 330, -183.3, 123.6 (Sept 27, 1976, May 11, 1997, May 25, 1998, Jun 2, 1998, May 30-31, 1999, Jun 1, 1999) | 15 - | 83, 34, 1, 260, -9.8, 8.6 (Apr 5, 1977, Apr 3, 1987, Nov 3, 1989, Aug 14, 1994) | 28 + | 93, 24, 6, 211, -70.3, 88.6 (Feb 26, 1978, Jan 13, 1982, Jan 14, 1982, Jan 15, 1982, Mar 26, 1996, Dec 29, 2009) |
| 3 - | 28, 18, 6, 330, -183.3, 123.6 (Nov 8, 1982, Sept 14, 1994, Dec 22-23, 2009, Dec 7, 2010) | 16 - | 84, 74, 41, 138, -141.7, 165.4 (Feb 22, 1977, Jul 24-25 1983, Jun 2, 1984, Jul 26, 1996) | 29 - | 103, 15, 316, 144, -1760.6, 1619.9 (Feb 22, 1977, Aug 17, 1994, Aug 23, 1994, May 8, 1995, Aug 11, 1995) |
| 4 + | 28, 18, 7, 329, -199.2, 134.9 (Nov 7, 1982, Sept 13, 1994, Dec 22-23, 2009, Dec 7, 2010) | 17 + | 85, 52, 2, 79, -65.0, 47.0 (Aug 3, 1988, May 29-30, 1994, Jul 1, 1998, Jun 23, 2003) | 30 - | 103, 39, 1, 21, -27.6, 23.2 (Jan 22, 1979, Jan 22, 1997, Jun 2, 1998, Aug 29, 2003) |
| 5 + | 28, 18, 7, 330, -199.2, 134.9 (Nov 8, 1982, Sept 9, 1994, Dec 23-24, 2009, Dec 8, 2010) | 18 + | 86, 48, 1, 319, -33.6, 25.3 (Dec 28, 1985, Jan 17, 1988, Oct 1, 2003, Dec 24, 2009) | 31 - | 105, 94, 1, 269, -22.0, 19.5 (Dec 12, 1981, May 6, 1984, Jan 4, 1998, Dec 8-9, 2010) |
| 6 + | 33, 25, 1, 106, -47.2, 22.8 (May 13, 1983, Jun 30, 1990, Jun 5, 1998, Aug 19, 2004) | 19 + | 86, 48, 1, 320, -33.6, 25.3 (Dec 29, 1985, Jan 18, 1988, Oct 2, 2003, Dec 2009) | 32 + | 109, 21, 3, 27, -59.9, 48.6 (Jul 23, 1982, Jan 20, 1997, Sept 6, 1999, Jun 20-21, 2006) |
| 7 - | 39, 37, 1, 21, -27.7, 22.6 (Jan 22, 1979, Jan 22, 1997, Jun 2, 1998, Aug 29, 2003) | 20 + | 86, 48, 2, 319, -60.0, 45.5 (Dec 29, 1985, Jan 17-18, 1988, Oct 1, 2003, Dec 23-24, 2009) | 33 - | 109, 27, 1, 28, -28.3, 17.9 (Oct 25, 1987, Jan 20, 1997, Jul 1, 1998, Jun 19, 2006) |
| 8 - | 43, 39, 1, 21, -29.3, 23.8 (Jan 22, 1979, Jan 22, 1997, May 2, 1998, Aug 29, 2003) | 21 + | 86, 48, 3, 318, -92.4, 66.0 (Dec 29, 1985, Jan 17, 1988, Oct 1, 2003, Dec 22-23, 2009) | 34 + | 116, 29, 4, 301, -159.4, 119.0 (Jan 16, 1975, Nov 7, 1985, Oct 17, 1988, Dec 4, 2004) |
| 9 - | 50, 9, 1, 38, -20.8, 17.1 (Jun 20, 1982, May 20, 1983, Feb 27, 1990, Mar 7, 1991) | 22 + | 86, 48, 4, 318, -116.7, 85.0 (Dec 29, 1985, Jan 17, 1988, Oct 2, 2003, Dec 22-24, 2009) | 35 + | 116, 33, 12, 315, -326.3, 233.4 (Dec 7, 1977, Dec 27, 1978, Nov 7, 1985, Jan 24, 2009) |
| 10 - | 55, 31, 1, 234, -17.2, 19.9 (Feb 22, 1977, Nov 8, 1984, Sept 8, 1994, Dec 6, 2010) | 23 + | 86, 48, 5, 317, -135.6, 101.9 (Dec 28-29, 1985, Jan 17-18, 1988, Oct 1-2, 2003, Dec 22-23, 2009) | 36 + | 119, 114, 3, 236, -75.0, 83.6 (Sept 5, 1982, Jul 30, 1983, Sept 14, 1994, Sept 2, 1998) |
| 11 + | 62, 1, 1, 293, -21.9, 22.2 (Dec 13, 1978, Jan 17, 1988, Sept 9, 2008, Dec 9, 2010) | 24 + | 89, 72, 2, 176, -49.9, 46.6 (Feb 6, 1979, Jan 17-18, 1988, Dec 23, 1996, Jan 6, 2001) | 37 + | 121, 91, 3, 131, -28.7, 22.7 (May 13-14, 1983, May 21, 1986, Apr 25, 19890, Jun 28 and 30, Jul 1, 1998) |
| 12 + | 62, 13, 2, 115, -46.2, 39.8 (Feb 18, 1978, Mar 23, 1991, Jun 2, 1998, Jun 14, 2010) | 25 - | 90, 71, 4, 96, -26.6, 55.8 (Jul 26-27, 1983, Sept 10, 1983, Jul 11, 1989, Aug 19, 2004) | 38 + | 121, 91, 4, 131, -36.9, 29.6 (May 13-14, 1983, May 21, 1986, Apr 25, 1989, Jun 29 – Jul 1, 1998) |
| 13 - | 76, 18, 3, 212, -86.2, 65.8 (Jul 31, 1979, Oct 4, 1987, Jul 25-26, 1996, Oct 9-10, 2001) | 26 + | 91, 44, 2, 165, -20.0, 15.7 (Aug 3, 1978, Jun 23, 1989, Aug 8, 1997, Aug 24, 2003) | 39 - | 124, 17, 2, 61, -11.5, 11.0 (Mar 15, 1983, Mar 14, 1999, Mar 5, 2001, Dec 8-9, 2010 |
| 14 - | 76, 18, 4, 211, -110.4, 84.1 (Jul 31, 1979, Oct 4, 1987, Jul 25, 1996, Oct 9-10, 2001) | 27 + | 92, 67, 3, 231, -205.9, 108.8 (Aug 31, 1987, Oct 12, 1990, Dec 8, 2000, Aug 30, 2007) | | |

A nonanticipative analog method for the long-range forecasting of heat/cold waves is used for the finding of teleconnections. The forecast method identifies the dependency between the current values of two meteorological variables and the future state of another variable (which may be identical to the first, as with air temperatures used here). The method was applied to the prediction of heat/cold waves for the several places around the world. The data include standard meteorological variables from 119 places around the world, as well as sea level (Aburatsu, Japan), monthly mean Darwin and Tahiti sea level pressures, SOI, equatorial SOI, sea surface temperature, and multivariate ENSO index (131 datasets in total). Every dataset is split into two samples, for learning and validation, respectively. Initially, the sum of the values at two different locations (minus corresponding climatological baselines) is calculated with lead-time from 14 to 365 days on summation interval of length from 1 to 365 days. Objective function defines the distribution based on two input datasets with appropriate lead-time and summation interval, which have maximum (or minimum) sum compared with the rest of data fourfold at least (with a minimum time difference of at least 30 days), when an extreme event occurs on the learning sample. Specific extreme events at Annaba (Algeria) were thus predicted on the validation sample based on rules referring to events in earlier years. Some heat waves are specifically predicted (up to 36.4% of all extremes). The methodology is 100% accurate if the signs of predicted and actual values are compared according to climatological baseline.

The most likely prospect for the further development of this work is quantum-computing system using D-Wave 2X, where qubits represent historical data in the forecast model.

Table 4. The forecasted heat waves at Annaba, validation sample from 2011 to 2014.

| Rule No. ($\overline{1,39}$) | Exact date of extreme value forecast | Forecasted sector of the month | Observed maximum value in forecasted sector of the month, $^0F$ | Climatological baseline, $^0F$ | SD, $^0F$ | Analysis |
|---|---|---|---|---|---|---|
| 16 28 | Feb 14-25 Feb 25, 2011 | The end of Feb 2011 | 55.1 Feb 24, 2011 | 53.1 | 4.0 | The same sign, difference is greater than climatological baseline |
| 16 9 2 | Feb 26-27 Mar 16 Mar 20, 2011 | The middle of Mar 2011 | 68.6 Mar 15, 2011 | 54.9 | 4.7 | Extreme value |
| 14 13 | Jun 10 Jun 11, 2011 | The middle of Jun 2011 | 82 Jun 18, 2011 | 72.2 | 3.7 | Extreme value |
| 13, 14 | Jul 1, 2011 | The middle of Jul 2011 | 84.8, 86.8, 85.7 Jul 11-13, 2011 | 76.9, 76.6, 76.1 | 3.5, 3.3, 3.3 | Extreme values |
| 3, 4 3-5 5 4 3-5 5 | Oct 8 Oct 9-11 Oct 12 Oct 27 Oct 28-30 Oct 31, 2011 | The end of Oct 2011 | 70.2 Oct 24, 2011 | 66.3 | 4.3 | The same sign, difference is greater than climatological baseline |
| 35 | Jan 6-10, 2012 | The middle of Jan 2012 | 52.3 Jan 11, 2012 | 52.0 | 3.7 | The same sign, difference is greater than climatological baseline |
| 32, 34 35 | Dec 2-5 Dec 16-25, 2012 | The beginning of Dec 2012 | 55.7 Dec 4, 2012 | 54.8 | 3.6 | The same sign, difference is greater than climatological baseline |
| 35 | Nov 4, 2013 | The beginning of Nov 2013 | 72.1 Nov 6, 2013 | 61.6 | 4.7 | Extreme value |
| 11 | Dec 23, 2013 | The middle of Dec 2013 | 54.7 Dec 20, 2013 | 53.1 | 3.6 | The same sign, difference is greater than climatological baseline |
| 28 | Jan 27, Feb 5, 2014 | The end of Jan, 2014 | 58.9 Jan 19, 2014 | 51.6 | 3.1 | Extreme value |
| 25 | Jul 19-20, 2014 | The middle of Jul, 2014 | 82.6, 83.8 Jul 19-20, 2014 | 76.9 76.8 | 2.5 2.7 | Extreme values |
| 6 | Sept 7, 2014 | The beginning of Sept 2014 | 83.8, 84.3 Sept 19-20, 2014 | 73.8 73.8 | 4.2 3.3 | This prediction is considered correct because of adjacency of the dates and significant lead-time 106 days. |
| 32 | Nov 11, 2014 | The middle of Nov, 2014 | 67.5, 66.1, 65.8, 64.5, 64.3, 68.3, 72.9, 73.9, 66.3 Nov 23 – Dec 1, 2014 | 56.8, 56.4, 56.6, 56.7, 56.3, 55.6, 55.2, 55.6, 55.3 | 4.6, 4.1, 3.8, 3.4, 3.8, 4.3, 3.5, 5.0, 4.2 | Extreme values |

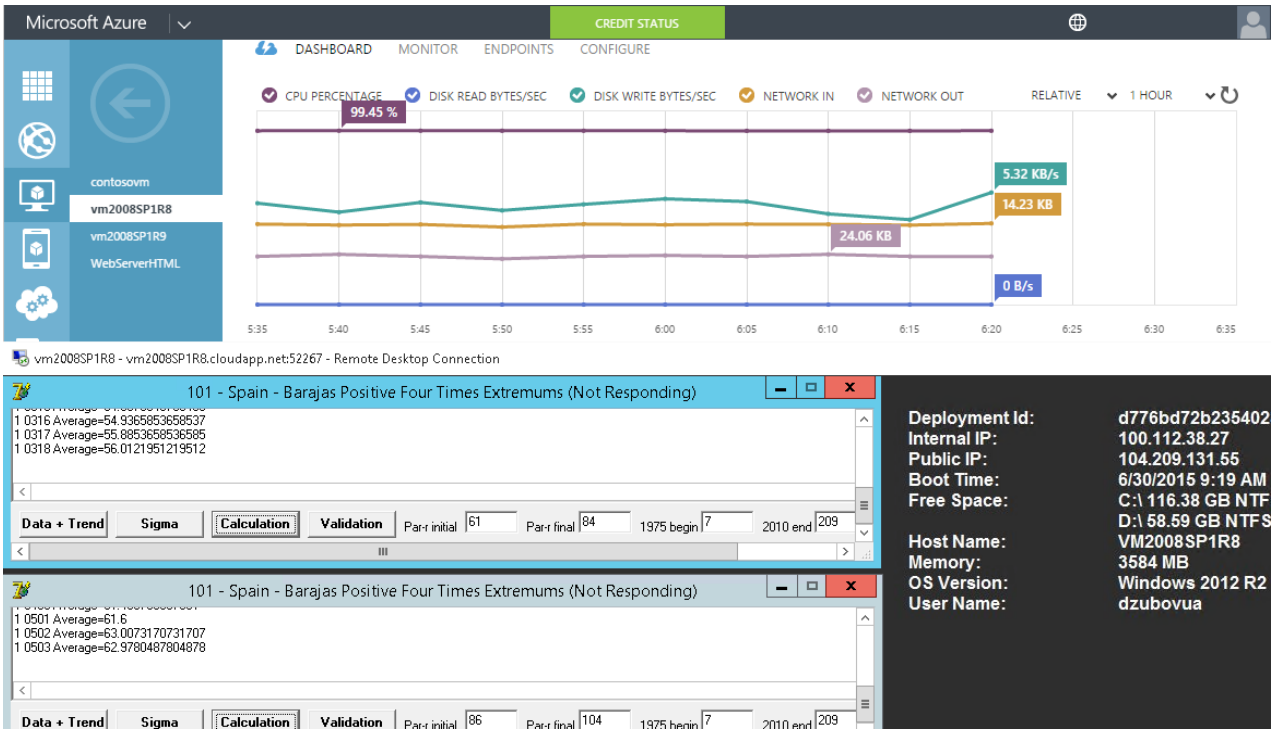

Fig. 1. Management portal of Windows Azure and VM with two Delphi desktop apps.

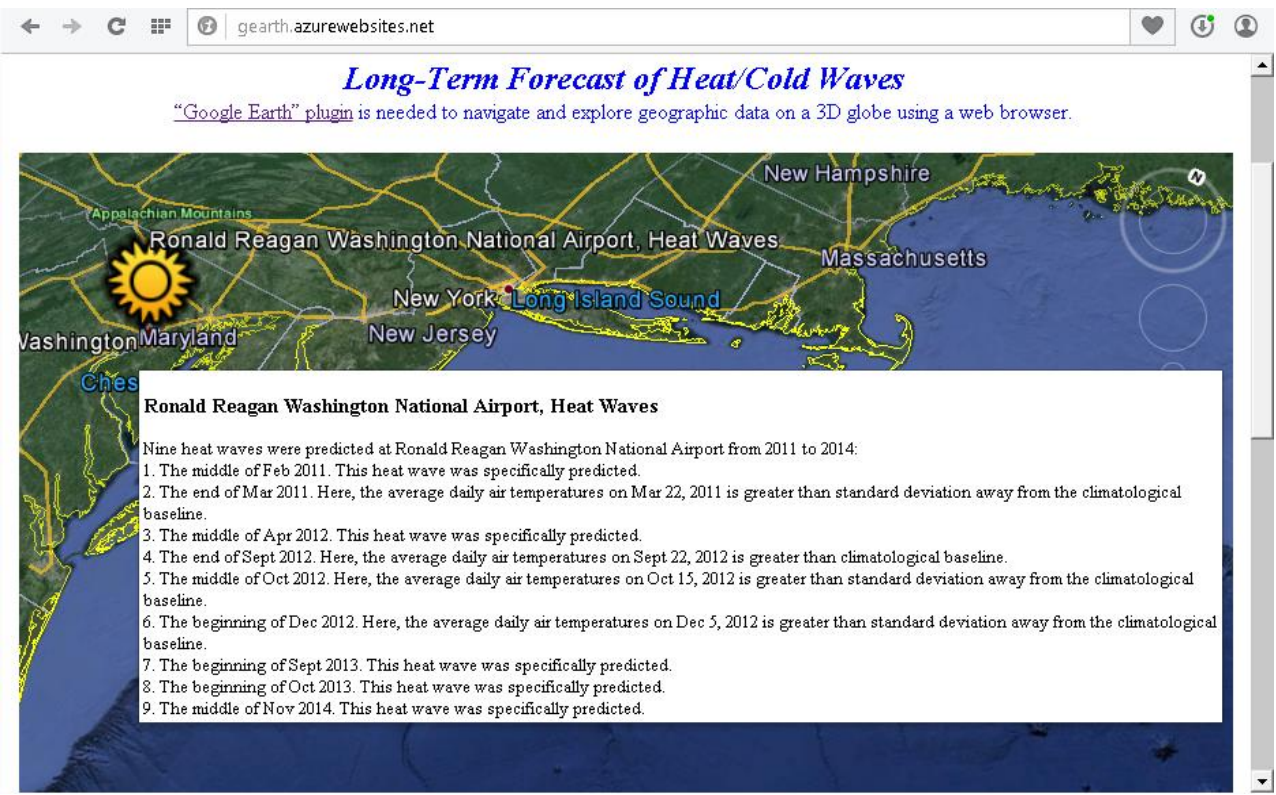

Fig. 2. Screenshot of the Google Earth Pro web-site for the visualization of heat/cold waves.

**Competing interests**
The author declares that he has no competing interests.

**Acknowledgements**

This work was jointly supported by the Microsoft Azure for Research program (Climate Data Initiative Award), Microsoft Educator Grant Program, University of Information Science and Technology "St. Paul the Apostle", and Tecnológico de Monterrey. The author wishes to express sincere appreciation and gratitude to all colleagues who took part in the research and supported in the preparation of this paper.